\documentclass[lettersize,journal]{IEEEtran}
\usepackage{amsmath,amsfonts}
\usepackage{algorithmic}
\usepackage{array}
\usepackage[caption=false,font=normalsize,labelfont=sf,textfont=sf]{subfig}
\usepackage{textcomp}
\usepackage{stfloats}
\usepackage{url}
\usepackage{verbatim}
\usepackage{graphicx}
\usepackage{xcolor}
\def\BibTeX{{\rm B\kern-.05em{\sc i\kern-.025em b}\kern-.08em
    T\kern-.1667em\lower.7ex\hbox{E}\kern-.125emX}}
\usepackage{balance}
\usepackage{framed}
\usepackage{float}
\usepackage{threeparttable} 
\usepackage{makecell}
\usepackage{multirow}
\usepackage{algorithm}
\begin{document}
\title{Power Optimization and Deep Learning for Channel Estimation of Active IRS-Aided IoT}
\author{Yan Wang, Feng Shu, Rongen Dong, Wei Gao, Qi Zhang, Jiajia Liu
\thanks{Manuscript created October, 2020;  This work was supported in part by the National Natural Science Foundation of China (Nos.U22A2002, and 62071234), the Hainan Province Science and Technology Special Fund (ZDKJ2021022), the Scientific Research Fund Project of Hainan University under Grant KYQD(ZR)-21008, and the Collaborative Innovation Center of Information Technology, Hainan University (XTCX2022XXC07). 
(Corresponding author: Feng Shu).}
\thanks{Yan Wang is with the School of Information and Communication Engineering, Hainan University, Haikou 570228, China (e-mail: yanwang@hainanu.edu.cn).}
\thanks{Feng Shu is with the School of Information and Communication Engineering and Collaborative Innovation Center of Information Technology, Hainan University, Haikou 570228, China, and also with the School of Electronic and Optical Engineering, Nanjing University of Science and Technology, Nanjing 210094, China (e-mail: shufeng0101@163.com).}
\thanks{Rongen Dong, Wei Gao, and Qi Zhang are with the School of Information and Communication Engineering, Hainan University, Haikou 570228, China (e-mail: dre2000@163.com; gaowei@hainanu.edu.cn; hdzhangqi0509@163.com).}
\thanks{Jiajia Liu is with the School of Cybersecurity, Northwestern Polytechnical University, Xi'an, Shaanxi, 710072, China (e-mail: liujiajia@nwpu.edu.cn).}
}


\maketitle

\begin{abstract}
In this paper, channel estimation of an active intelligent reflecting surface (IRS) aided uplink Internet of Things (IoT) network is investigated. Firstly, the least square (LS) estimators for the direct channel and the cascaded channel are presented, respectively. The corresponding mean square errors (MSE) of channel estimators are derived. Subsequently, in order to evaluate the influence of adjusting the transmit power at the IoT devices or the reflected power at the active IRS on Sum-MSE performance, two situations are considered. In the first case, under the total power sum constraint of the IoT devices and active IRS, the closed-form expression of the optimal power allocation factor is derived. In the second case, when the transmit power at the IoT devices is fixed, there exists an optimal reflective power at active IRS. To further improve the estimation performance, the convolutional neural network (CNN)-based direct channel estimation (CDCE) algorithm and the CNN-based cascaded channel estimation (CCCE) algorithm are designed. Finally, simulation results demonstrate the existence of an optimal power allocation strategy that minimizes the Sum-MSE, and further validate the superiority of the proposed CDCE / CCCE algorithms over their respective traditional LS and minimum mean square error (MMSE) baselines.
\end{abstract}


\section{Introduction}

With the unceasing advancement of wireless communication technology, the sixth generation (6G) networks are poised to facilitate human-centric immersive interactive experiences, while simultaneously guaranteeing the efficient and reliable functioning of Internet of Things (IoT) scenarios \cite{NaSoft} \cite{GuoEnabling}.
The strengthened massive Machine Type Communication (mMTC) emerges as a critical application case in 6G networks, broadening the connectivity spectrum to encompass diverse devices across scenarios such as smart cities, intelligent living, smart transportation, precision agriculture, and advanced manufacturing \cite{GaoEnergy}. This evolution is anticipated to offer robust support for the proliferation of large-scale IoT devices and their emerging applications.
Some researchers predict that by 2040, IoT terminals will show an explosive growth of hundreds of billions and more than 98\% of connections \cite{WangOn}. In other words, the widespread adoption of IoT dominates the development direction of communication technology to a certain extent.

Concurrently, traditional IoT networks encounter several noteworthy challenges \cite{JiangJoint} \cite{ZhengSimultaneous}. For instance, IoT nodes / users are typically dispersed across vast geographical areas, resulting in some users experiencing shadow fading due to obstructions in their surroundings \cite{YangIntelligent}. This shadowing effect can compromise the network's coverage and overall performance. In addition, the constrained battery capacity of IoT devices, stemming from their physical size and cost limitations, which poses a significant challenge, inherently restricting their ability to upload information \cite{ChenActive}.
To effectively improve the coverage as well as energy efficiency in IoT networks, intelligent reflecting surface (IRS) \cite{ShuEnhanced} \cite{ShuBeamforming} technology is gradually receiving substantial attention. Owing to its unparalleled configurability and intelligent attributes, the IRS is anticipated to occupy a pivotal position in IoT networks, thereby propelling the evolution of wireless networks towards a more efficient and environmentally sustainable trajectory \cite{YuDesign}.

IRS-assisted IoT networks have recently received increasing research.
An IRS-aided unmanned aerial vehicle (UAV) communication network was proposed in \cite{MamaghaniAerial}, which was used to disseminate confidential information from an access point (AP) to terrestrial devices in an IoT network.
Subsequently, the power allocation of AP, the beamforming of IRS, and the trajectory of the UAV were jointly designed. 
Besides, the authors of \cite{MahmoudIntelligent} discussed the performance analysis of IRS-aided UAV communication in IoT networks, derived closed-form expressions for the achievable symbol error rate (SER), outage probability, and ergodic capacity, respectively.
Furthermore, an IRS-assisted wireless powered sensor network (WPSN) was studied in \cite{ChuZhuIntelligent}. 
Specifically, the power station provided wireless energy to multiple IoT devices so that they could deliver their respective messages to the AP. 
To maximize the total throughput of IRS assisted WPSN, the phase shift matrix and transmission time allocation were jointly optimized.
Similarly, the authors of \cite{ZhuResource} investigated a new transmission strategy for IRS empowered IoT systems. 
In addition, the power allocation (PA) ratio in \cite{ZhuResource} was considered to maximize the total throughput.
The authors of \cite{ChuXiaoIntelligent} deployed the IRS in a mobile edge computing (MEC) network to ameliorate the computational performance, by flexibly manipulating the phase shift of each reflective element.

In summary, the existing research on IRS-assisted IoT networks encompasses notable advancements in the enhancement of network capacity \cite{DiaoJoint}, the maximization of received signal-to-noise ratio (SNR) \cite{ShuThree}, the optimization of energy efficiency \cite{WangWireless-Powered} and spectral efficiency, as well as the exploration of IRS-aided physical layer security. 
However, the diverse performance enhancements attained by the IRS are highly reliant on precise channel state information (CSI) \cite{GuanJointPower} \cite{GuanAnchor-Assisted}. Consequently, acquiring an accurate CSI for an IRS-aided IoT network is of paramount importance \cite{ZhengASurvey} \cite{ZhengIntelligent}.

\subsection{Prior Works}

Recently, some channel estimation (CE) schemes have been designed for IRS-assisted wireless networks \cite{WuChannel,YouWireless,ZhouChannel}. 
In general, the CSI for direct channels (DCs) can be obtained through some traditional classical algorithms \cite{ShuPilot}.
Some researchers have pointed out that the cascaded CE issue can be addressed through algorithms, such as least square (LS), or minimum mean square error (MMSE). 
It is worth noting that the construction of cascaded channels (CCs) involves the product of the number of base station (BS) antennas and the number of IRS elements.
Specifically, compared to conventional large-scale MIMO CE schemes that do not include IRS, IRS assisted communication systems face the challenge of rapidly increasing pilot overhead in the cascaded CE process.
To effectively mitigate the above mentioned pilot overhead, the authors of \cite{WangChannel} designed a CE algorithm based on a three-phase pilot design, which was customized for IRS-assisted uplink multiuser communication systems.
Furthermore, a low pilot overhead strategy for MIMO uplink transmission was explored in \cite{RoshdyA}.
This study adopted a two-stage CE framework: in the first phase, an efficient pilot transmission scheme in time division duplex mode was used to focus on CE between BS and IRS.
In phase II, the BS-IRS channel information obtained in the first stage was utilized to further derive the channel status between IRS and IoT devices. This method not only improved the efficiency of CE, but also promoted the practical application of IRS assisted communication systems in resource limited scenarios.

In addition, in \cite{GuoJoint}, researchers delved into the joint activity detection and CE issue in distributed and centralized IRS-aided IoT networks. This study successfully achieved the MMSE estimation of effective channel coefficients by introducing the generalized approximate message passing (GAMP) algorithm.
It is worth noting that the CE task in integrated sensing and communication (ISAC) systems is facing unprecedented challenges due to the lack of signal processing capability in passive IRS and the existence mutual interference between sensing signals and communication signals.
To address this challenge, the author of \cite{LiuDeep-Learning} innovatively proposed a three-stage method that ingeniously decomposed the complex estimation problem into three sub problems: DC, reflection communication channel, and reflection sensing channel. Each sub problem was overcome one by one, effectively reducing the complexity and improving the accuracy of the estimation.
In order to further overcome the constraint of CE, a received power-based CE scheme was presented in \cite{JungRSS-Based}. This scheme, with its advantages of simple implementation and easy deployment, provided a novel and practical approach for CE.

In IRS-aided wireless networks, some deep learning (DL) methods have been widely adopted to enhance the performance of CE.
For instance, in \cite{LiuDeepResidual}, the authors regarded the CE process as a denoising task and used deep residual learning (DRL) techniques to implicitly learn and removed residual noise from the observed signal, thereby achieving the goal of accurately recovering channel coefficients from noisy pilot observations.
Besides, distributed machine learning (ML) techniques were used to achieve reliable downlink CE \cite{DaiDistributed}. Particularly, the authors of \cite{ShenDeep} formulated CE as an image super-resolution problem, and leveraged convolutional neural network (CNN) to propose an image super-resolution network. This network utilized the backpropagation (BP) algorithm to iteratively adjust its trainable parameters until the model was fully trained.

There are also several solutions on how to obtain CSI under the Rayleigh fading channel model. 
First of all, a dual-link pilot transmission algorithm was presented in \cite{HuDaiTwo-Timescale}, in which the BS was responsible for transmitting downlink pilot signals and receiving uplink pilot signals reflected by IRS.
Subsequently, a coordinate descent-based scheme to recover the BS-RIS channel was presented. 
Simulation results showed that the proposed two-timescale CE scheme could accomplish high-precision CE while reducing pilot overhead.
Furthermore, two CE algorithms were proposed in \cite{WeiChannel} for the channel between BS and IRS, as well as users and IRS.
The first algorithm used alternating least squares (ALS), while the other algorithm utilized vector approximation message passing (VAMP) technique, which iteratively reconstructed two unknown channel matrices from initially estimated vectors.
Through comparative analysis, it was found that the VAMP-based algorithm had lower complexity over the ALS-based algorithm.
Simulation results further verified that the performance of these two algorithms was quite similar and significantly better than the benchmark scheme.
Moreover, in \cite{AwaisDeep}, researchers designed a modified DRL U-shaped architecture aimed at capturing and extracting the spatial characteristics of the IRS channel to achieve efficient signal denoising.

\subsection{Our Contributions}

To the best of our knowledge, this is the first attempt to explore CE methods for active IRS-aided uplink IoT networks. The main contributions of this paper are
summarized as follows:

\begin{enumerate}
   \item  Firstly, an active IRS-aided uplink IoT network is constructed. Unlike previous studies that often overlooked the estimation of DCs, the DCs and CCs are perfectly separated by adopting a clever pilot pattern. Subsequently, LS channel estimators for the DC and the CC are proposed, respectively. Correspondingly, the closed-form expressions of the MSE for both DC and CC are derived.
  \item In order to evaluate the impact of adjusting the transmit power at the IoT device or the reflected power at active IRS on the Sum-MSE, two cases are considered. In the first case, an optimal PA strategy that minimizes the Sum-MSE is given under the total power sum constraint of the IoT device and active IRS. Specifically, a closed-form expression of the optimal power allocation factor (PAF) is derived. In the second case, when the transmit power at the IoT device is fixed, there exists an optimal limited reflective power at active IRS. Beyond which an increase in reflected power at active IRS will adversely impact the Sum-MSE performance.
  \item To leverage the powerful learning and nonlinear modeling capabilities of neural networks for high-dimensional CE, two uplink CE networks have been proposed, namely the CNN-based direct channel estimation (CDCE) algorithm and the CNN-based cascaded channel estimation (CCCE) algorithm. Drawing on the advantages of CNN in feature extraction, the proposed two CE methods can further improve the estimation accuracy. Simulation results validate the superiority of the proposed CDCE / CCCE network over the corresponding LS and MMSE benchmark schemes, and their computational complexity is also analyzed.

\end{enumerate}

\subsection{Organization and Notation}

The remainder of this paper is structured as follows. In Section II, the system model of an active IRS-aided uplink IoT network is established. In Section III, we first propose the LS channel estimator and MMSE estimator, then, the optimal PA strategy is investigated to minimize Sum-MSE. In Section IV, the CDCE algorithm and the CCCE algorithm are proposed, and the computational complexity of the two proposed networks is also given. Simulation results and conclusions are provided in Section V and Section VI, respectively.

Notations: Throughout the paper, vectors and matrices are denoted by boldface lower case and upper case letters, respectively. Lower case letters are employed to represent scalars, while $(\cdot)^{H}$ stands for conjugate and transpose operation. 
The sign $(\cdot)^{T}$ denotes the transpose operation.
The signs $\|\cdot\|_2$ and $\|\cdot\|_F$ stand for the 2-norm and $F$-norm, respectively. 
The notation $\mathbb{E}\{\cdot\}$ denotes the expectation operation. $\hat{[\;]}$ represents the estimation operation.

\section{System model}
\begin{figure}[htbp]
\centering
\includegraphics[width=3.5in]{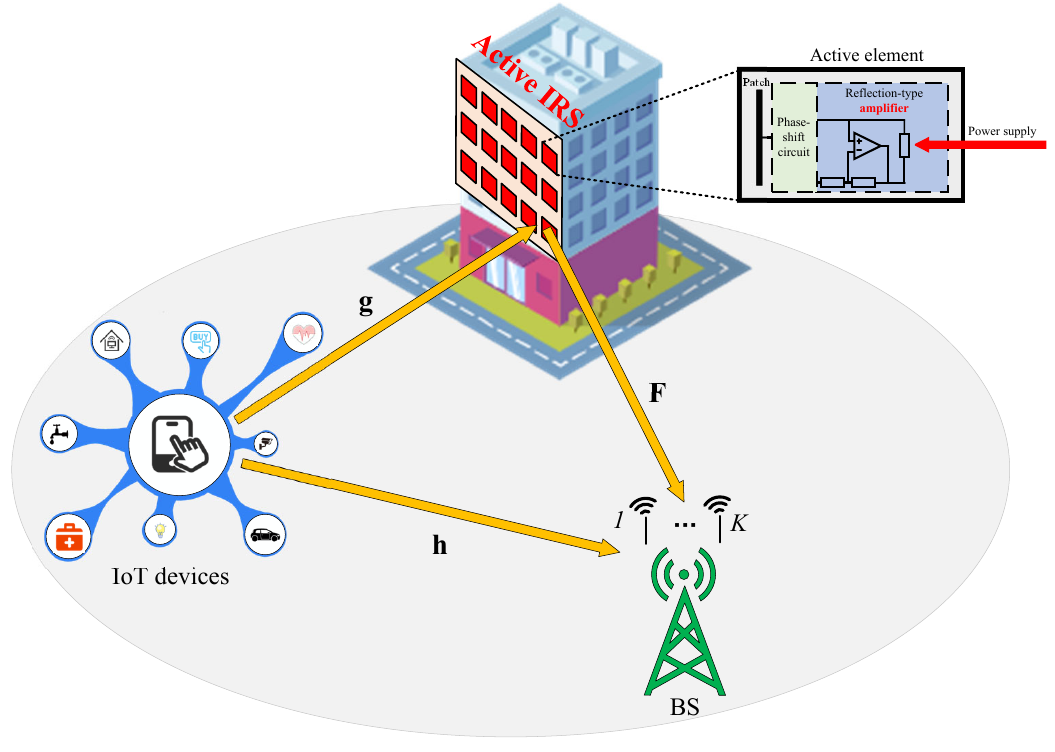}\\
\caption{A system model of an active IRS-aided uplink IoT network.}\label{Visio-channel-estimation}
\end{figure}

The active IRS-assisted uplink IoT network is shown in Fig.~\ref{Visio-channel-estimation}, where the IoT device is configured with single antenna, the BS is equipped with $K$ antennas, and IRS employs $N$ active reflecting elements.
It is worth noting that compared to passive IRS that can only passively reflect signals, active IRS is unique due to the integration of reflection-type amplifiers inside its elements. In other words, active IRS not only actively reflects signals, but also has the ability to amplify signals.
Assuming that IoT devices to active IRS, IoT devices to BS, and active IRS to BS are all Rayleigh fading channels.
It is worth noting that BS can estimate the channel characteristics between a large number of antennas through a small number of pilot signals in the uplink. Based on channel reciprocity, it can effectively predict or reconstruct the CSI of the downlink.

%
%

The transmitted signal at IoT device is given as follows
\begin{align}
s=\sqrt{\beta P_t}x,
\end{align}
where $x$ is the transmit symbol with unit energy, $P_t$ represents the power sum of the IoT devices and active IRS, and $\beta$ denotes the PAF.

The received signal at active IRS is written as
\begin{align}
\mathbf{y}_{\text{IRS}}=\sqrt{\beta P_t}\mathbf{\Theta}\mathbf{g}x+\mathbf{\Theta}\mathbf{w}_i,
\end{align}
where $\mathbf{\Theta}\in\mathbb{C}^{N\times N}$ stands for the phase shift of the active IRS, and $\mathbf{g}\in\mathbb{C}^{N\times 1}$ denotes the channel between IoT devices and active IRS. 
$\mathbf{w}_i$ represents the additive white Gaussian noise (AWGN) at active IRS with $\mathbf{w}_i\in \mathbb{C}^{N\times 1}\sim \mathcal{CN}(\mathbf{0},\sigma_i^2\mathbf{I}_N)$.

The received signal at BS can be modeled as follows
\begin{align}\label{y1new}
\mathbf{y}=\sqrt{\beta P_t}\mathbf{h}x+\sqrt{\beta P_t}\mathbf{F}\mathbf{\Theta}\mathbf{g}x
+\mathbf{F}\mathbf{\Theta}\mathbf{w}_i
+\mathbf{w}_b,
\end{align}
where $\mathbf{h}\in\mathbb{C}^{K\times 1}$ and $\mathbf{F}\in\mathbb{C}^{K\times N}$ represent the channels between IoT devices and BS, as well as the channels between active IRS and BS, respectively. 
$\mathbf{w}_b$ is the AWGN at BS with $\mathbf{w}_b\in \mathbb{C}^{K\times 1}\sim \mathcal{CN}(\mathbf{0},\sigma_b^2\mathbf{I}_K)$.

In (\ref{y1new}), $\mathbf{\Theta}\mathbf{g}$ can be rewritten as $\mathbf{\Theta}\mathbf{g}=\text{diag}\{\mathbf{g}\}\boldsymbol{\theta}
=\mathbf{G}\boldsymbol{\theta}$, where $\mathbf{\Theta}=\text{diag}\{\boldsymbol{\theta}\}$.
Therefore, the CC can be represented as $\mathbf{H}_{biu}=\mathbf{F}\mathbf{G}=\mathbf{F}\text{diag}\{\mathbf{g}\}\in \mathbb{C}^{K\times N}$, (\ref{y1new}) can be represented as follows
\begin{align}\label{ysqrt}
\mathbf{y}
&=\sqrt{\beta P_t}\mathbf{h}x+\sqrt{\beta P_t}\mathbf{F}\mathbf{G}
\boldsymbol{\theta}x
+\mathbf{F}\text{diag}\{\mathbf{w}_i\}
\boldsymbol{\theta}+\mathbf{w}_b\nonumber\\
&=\sqrt{\beta P_t}\mathbf{h}x+\sqrt{\beta P_t}\mathbf{H}_{biu}
\boldsymbol{\theta}x
+\mathbf{F}\mathbf{W}_i
\boldsymbol{\theta}+\mathbf{w}_b.
\end{align}

For active IRS, let us define $\boldsymbol{\theta}=\rho \boldsymbol{\tilde{\theta}}\in\mathbb{C}^{N\times 1}$, where $\rho= \|\boldsymbol{\theta}\|_2/\sqrt{N}$ and $\|\boldsymbol{\tilde{\theta}}\|_2=\sqrt{N}$.
Then, (\ref{ysqrt}) can be expressed as
%
%
%
%
\begin{align}
\mathbf{y}=\sqrt{\beta P_t}\mathbf{h}x
+\rho\sqrt{\beta P_t}\mathbf{H}_{biu}
\boldsymbol{\tilde{\theta}}x
+\rho\mathbf{F}\mathbf{W}_i
\boldsymbol{\tilde{\theta}}+\mathbf{w}_b.
\end{align}

Correspondingly, the power reflected by active IRS is
\begin{align}\label{PIRS}
P_{\text{IRS}}
&=\mathbb{E}\{\mathbf{y}_{\text{IRS}}^H
\mathbf{y}_{\text{IRS}}\}\\
&=\rho^2\beta P_t\|\mathbf{G}\boldsymbol{\tilde{\theta}}\|_2^2
+\rho^2\sigma_i^2
=(1-\beta)P_t.\nonumber
\end{align}

From (\ref{PIRS}), $\rho$ can be formulated as
\begin{align}\label{rho}
\rho=\sqrt{\frac{(1-\beta)P_t}
{\beta P_t\|\mathbf{G}\boldsymbol{\tilde{\theta}}\|_2^2
+\sigma_i^2}}.
\end{align}

\section{Proposed LS Channel Estimator and Power Optimization}

In this section, first of all, LS channel estimators are proposed for direct and cascaded channels, respectively. Subsequently, we delve into the PA strategy aimed at minimizing the Sum-MSE.

\subsection{Proposed LS Channel Estimator}

\begin{figure}[htbp]
\centering
\includegraphics[width=3.5in]{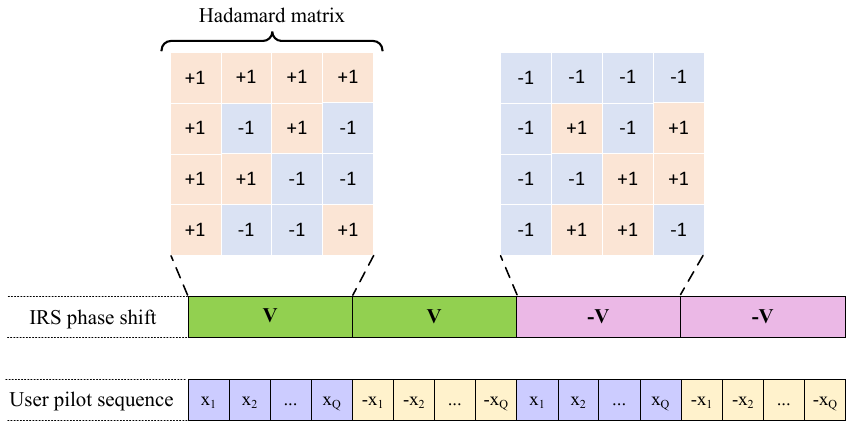}\\
\caption{Pilot pattern.}\label{Visio-Pilot}
\end{figure}

To facilitate subsequent CE, we aim to segregate the DC and the CC. Drawing inspiration from \cite{SunZPilot}, we adopt a specific pilot pattern depicted in Fig.~\ref{Visio-Pilot}, where a Hadamard matrix is employed as an illustrative example of this pattern.

During the first pilot sequence period, we have
\begin{align}
\mathbf{Y}_1=&[\mathbf{y}_1,\mathbf{y}_2,\ldots,\mathbf{y}_{Q}]
\\
=&\sqrt{\beta P_t}\mathbf{h}\mathbf{x}^T
+\rho\sqrt{\beta P_t}\mathbf{H}_{biu}\mathbf{V}\mathbf{X}
+\rho\mathbf{F}\mathbf{V} \mathbf{W}_{i1} +\mathbf{W}_{b1},\nonumber
\end{align}
where the pilot sequence $\mathbf{x}=[x_1, x_2, \cdots, x_{Q}]^T\in \mathbb{C}^{Q\times 1}$ of $Q$ symbols is sent by the IoT devices, and $\mathbf{X}=\text{diag}\{\mathbf{x}\}\in \mathbb{C}^{Q\times Q}$.
Moreover, each symbol comprising the pilot sequence is associated with a distinct phase configuration of the active IRS $\boldsymbol{\theta}_i$, and $i=1, 2, \cdots, Q$. Furthermore, $\mathbf{V}=[\boldsymbol{\tilde{\theta}}_1, \boldsymbol{\tilde{\theta}}_2, \cdots, \boldsymbol{\tilde{\theta}}_{Q}]\in \mathbb{C}^{N\times Q}$, and $\mathbf{W}_{b1}=[\mathbf{w}_{b1}, \mathbf{w}_{b2}, \cdots, \mathbf{w}_{bQ}]\in \mathbb{C}^{K\times Q}$.

Similarly, in the last three pilot sequences, the three received signal are given as follows
\begin{align}
\mathbf{Y}_2=-\sqrt{\beta P_t}\mathbf{h}\mathbf{x}^T
-\rho\sqrt{\beta P_t}\mathbf{H}_{biu}\mathbf{V}\mathbf{X}
+\rho\mathbf{F}\mathbf{V}\mathbf{W}_{i2} +\mathbf{W}_{b2},
\end{align}
\begin{align}
\mathbf{Y}_3=\sqrt{\beta P_t}\mathbf{h}\mathbf{x}^T
-\rho\sqrt{\beta P_t}\mathbf{H}_{biu}\mathbf{V}\mathbf{X}
-\rho\mathbf{F}\mathbf{V}\mathbf{W}_{i3} +\mathbf{W}_{b3},
\end{align}
\begin{align}
\mathbf{Y}_4=-\sqrt{\beta P_t}\mathbf{h}\mathbf{x}^T
+\rho\sqrt{\beta P_t}\mathbf{H}_{biu}\mathbf{V}\mathbf{X}
-\rho\mathbf{F}\mathbf{V}\mathbf{W}_{i4} +\mathbf{W}_{b4}.
\end{align}

Owing to the symmetric nature of the aforementioned four formulas, the following individual formulas for the cascaded and direct channels have been readily derived. The details are as follows:
\begin{align}\label{Y1tilde}
\tilde{\mathbf{Y}}_1=&\mathbf{Y}_1-\mathbf{Y}_2+\mathbf{Y}_3
-\mathbf{Y}_4\\
=&4\sqrt{\beta P_t}
\mathbf{h}\mathbf{x}^T\nonumber\\ 
&+\rho(\mathbf{F}\mathbf{V}\mathbf{W}_{i1}
-\mathbf{F}\mathbf{V}\mathbf{W}_{i2}
+\mathbf{F}\mathbf{V}\mathbf{W}_{i3}
-\mathbf{F}\mathbf{V}\mathbf{W}_{i4})\nonumber\\ 
&+\mathbf{W}_{b1}-\mathbf{W}_{b2}
+\mathbf{W}_{b3}-\mathbf{W}_{b4},\nonumber
\end{align}
\begin{align}\label{Y2tilde}
\tilde{\mathbf{Y}}_2=&\mathbf{Y}_1-\mathbf{Y}_2-\mathbf{Y}_3
+\mathbf{Y}_4\\
=&4\rho\sqrt{\beta P_t}\mathbf{H}_{biu}\mathbf{V}\mathbf{X}\nonumber\\ 
&+\rho(\mathbf{F}\mathbf{V}\mathbf{W}_{i1}
-\mathbf{F}\mathbf{V}\mathbf{W}_{i2}
-\mathbf{F}\mathbf{V}\mathbf{W}_{i3} 
+\mathbf{F}\mathbf{V}\mathbf{W}_{i4})\nonumber\\ 
&+\mathbf{W}_{b1}-\mathbf{W}_{b2}-\mathbf{W}_{b3}
+\mathbf{W}_{b4},\nonumber
\end{align}
where $Q\geq N$ is used to ensure the existence of the right pseudo-inverse of $\mathbf{V}$. To minimize overheads, it is assumed that $Q=N$. Subsequently, by right multiplying $\mathbf{x}^*$ on both sides of (\ref{Y1tilde}) at the same time, we can yield
\begin{align}
\tilde{\mathbf{Y}}_1\mathbf{x}^{*}=&4\sqrt{\beta P_t}\mathbf{h}\mathbf{x}^T\mathbf{x}^{*}\nonumber\\
+&\rho(\mathbf{F}\mathbf{V}\mathbf{W}_{i1}
-\mathbf{F}\mathbf{V}\mathbf{W}_{i2}
+\mathbf{F}\mathbf{V}\mathbf{W}_{i3} 
-\mathbf{F}\mathbf{V}\mathbf{W}_{i4})\mathbf{x}^{*}\nonumber\\
+&(\mathbf{W}_{b1}-\mathbf{W}_{b2}
+\mathbf{W}_{b3}-\mathbf{W}_{b4})\mathbf{x}^{*},
\end{align}
then, the LS estimator of $\mathbf{h}$ is
\begin{align}
\hat{\mathbf{h}}=\frac{\tilde{\mathbf{Y}}_1\mathbf{x}^{*}}
{4\sqrt{\beta P_t}\mathbf{x}^T\mathbf{x}^{*}}.
\end{align}
Furthermore, the MMSE estimator for the DC can be formulated as
\begin{align}
\hat{\mathbf{h}}_{\text{MMSE}}=
\mathbf{W}_{\text{MMSE}}\hat{\mathbf{h}}.
\end{align}
Subsequently, the MMSE estimator derives an estimate based on $\mathbf{W}$ that minimizes the MSE in the following equation
\begin{align}\label{DirectvarepsilonMMSE}
\varepsilon_1^{\prime}=\mathbb{E}\{\|\mathbf{h}
-\mathbf{W}_{\text{MMSE}}\hat{\mathbf{h}}\|_F^2\}.
\end{align}
By resolving the problem posed in equation (\ref{DirectvarepsilonMMSE}), the following results can be obtained:
\begin{align}
\mathbf{W}_{\text{MMSE}}=
\mathbf{R}_{\mathbf{h}\hat{\mathbf{h}}}
(\mathbf{R}_{\hat{\mathbf{h}}\hat{\mathbf{h}}}
+\frac{\sigma_b^2}{\sigma_x^2}\mathbf{I})^{-1},
\end{align}
where $\mathbf{R}_{\mathbf{h}\hat{\mathbf{h}}}=
\mathbb{E}\{\mathbf{h}\hat{\mathbf{h}}^H\}$ represents the cross-correlation matrix (CCM) between the true DC and the LS CE, while $\mathbf{R}_{\hat{\mathbf{h}}\hat{\mathbf{h}}}=
\mathbb{E}\{\hat{\mathbf{h}}\hat{\mathbf{h}}^H\}$ denotes the autocorrelation matrix (AM) of the LS CE.

By performing the vec operation on both sides of equation (\ref{Y2tilde}), and according to $\text{vec}(\mathbf{A}\mathbf{C})=(\mathbf{C}^T\otimes\mathbf{I})
\text{vec}(\mathbf{A})$, the following expression is derived:
\begin{align}
&\text{vec}(\tilde{\mathbf{Y}}_2)
=4\rho\sqrt{\beta P_t}
(\underbrace{(\mathbf{V}\mathbf{X})^T\otimes\mathbf{I}_K}_{\mathbf{A}_s})
\text{vec}(\mathbf{H}_{biu})\nonumber\\
&+\rho\text{vec}(\mathbf{F}\mathbf{V}\mathbf{W}_{i1}
-\mathbf{F}\mathbf{V}\mathbf{W}_{i2}
-\mathbf{F}\mathbf{V}\mathbf{W}_{i3} 
+\mathbf{F}\mathbf{V}\mathbf{W}_{i4})\nonumber\\
&+\text{vec}(\mathbf{W}_{b1}-\mathbf{W}_{b2}
-\mathbf{W}_{b3}
+\mathbf{W}_{b4}),
\end{align}
which subsequently yields the LS estimation of $\mathbf{H}_{biu}$ as follows
\begin{align}\label{CascadeLS}
\mathrm{vec}(\hat{\mathbf{H}}_{biu})
=\frac{\mathbf{A}_s^{-1}\mathrm{vec}(\tilde{\mathbf{Y}}_2)}
{4\rho\sqrt{\beta P_t}}.
\end{align}
Similarly, the MMSE estimator of the CC is
\begin{align}
\hat{\mathbf{h}}^{biu}_{\text{MMSE}}=
\mathbf{R}_{\mathbf{h}_{biu}\hat{\mathbf{h}}_{biu}}
(\mathbf{R}_{\hat{\mathbf{h}}_{biu}
\hat{\mathbf{h}}_{biu}}
+\frac{\sigma_b^2}{\sigma_x^2}\mathbf{I})^{-1}
\hat{\mathbf{h}}_{biu},
\end{align}
where $\hat{\mathbf{h}}_{biu}
=\mathrm{vec}(\hat{\mathbf{H}}_{biu})$. In addition, the CCM between the true CC and its corresponding LS CE is denoted as $\mathbf{R}_{\mathbf{h}_{biu}
\hat{\mathbf{h}}_{biu}}=
\mathbb{E}\{\mathbf{h}_{biu}
\hat{\mathbf{h}}_{biu}^H\}$, and $\mathbf{R}_{\hat{\mathbf{h}}_{biu}\hat{\mathbf{h}}_{biu}}=
\mathbb{E}\{\hat{\mathbf{h}}_{biu}\hat{\mathbf{h}}_{biu}^H\}$ stands for the AM of the LS CE of the CC.

Given $\mathbb{E}\{\mathbf{x}^T\mathbf{x}^*\}=N$, the MSE of estimating $\mathbf{h}$ is 
\begin{align}
\varepsilon_1&=\frac{1}{4N}
\mathbb{E}\{\|\hat{\mathbf{h}}-\mathbf{h}\|_F^2\}\nonumber\\
&=\frac{1}{4N}\mathbb{E}\bigg\{\bigg\|
\frac{\tilde{\mathbf{Y}}_1\mathbf{x}^{*}}
{4\sqrt{\beta P_t}\mathbf{x}^T\mathbf{x}^{*}}
-\mathbf{h}\bigg\|_F^2\bigg\}\nonumber\\
&=\frac{\rho^2\sigma_i^2\|\mathbf{F}\mathbf{V}\|_F^2
+\sigma_b^2}
{16N^2\beta P_t}.
\end{align}

Similarly, the MSE for estimating $\mathbf{H}_{biu}$ can be expressed as
\begin{align}
\varepsilon_2&=\frac{1}{4N}\mathbb{E}\{\|\mathrm{vec}
(\hat{\mathbf{H}}_{biu}
)-\mathrm{vec}(\mathbf{H}_{biu})\|_F^2\}\nonumber\\
&=\frac{\rho^2\sigma_i^2\|\mathbf{B}\|^2_F
+\sigma_b^2\|\mathbf{A}_s^{-1}\|^2_F}{16N\rho^2\beta P_t},
\end{align}
where $\mathbf{B}=\mathbf{A}_s^{-1}
\big(\mathbf{I}_N\otimes
(\mathbf{F}\mathbf{V})\big)$.

The Sum-MSE can be obtained directly as follows
\begin{align}\label{summse}
\varepsilon&=\varepsilon_1+\varepsilon_2\\
&=\frac{\rho^4\sigma_i^2\|\mathbf{F}\mathbf{V}\|_F^2
+\rho^2\sigma_b^2
+N\rho^2\sigma_i^2\|\mathbf{B}\|^2_F
+N\sigma_b^2\|\mathbf{A}_s^{-1}\|^2_F}{16N^2\rho^2\beta P_t}.\nonumber
\end{align}

%
%

\subsection{Power Optimization of minimizing Sum-MSE}

In this subsection, two scenarios are considered, which are:

\subsubsection{When the total power sum of the IoT devices and active IRS is limited} there exists an optimal PA strategy. 

To find the optimal PAF $\beta$, we first convert (\ref{summse}) to a function of $\beta$, i.e., substituting (\ref{rho}) into (\ref{summse}) yields
\begin{align}\label{varepsilon-beta}
\varepsilon(\beta)=\frac{a_1\beta^2+a_2\beta+a_3}
{a_4\beta^3+a_5\beta^2+a_6\beta},
\end{align}
%
where
\begin{align}
a_1&=b_2P_t^2-b_1(\sigma_b^2+b_3)P_t+b_1^2b_4,\nonumber\\
a_2&=-2b_2P_t^2+(\sigma_b^2+b_3)(b_1-\sigma_i^2)P_t
+2b_1b_4\sigma_i^2,\nonumber\\
a_3&=b_2P_t^2+(\sigma_b^2+b_3)\sigma_i^2P_t+b_4\sigma_i^4,\nonumber\\
a_4&=-b_1b_5P_t,\nonumber\\
a_5&=(b_1-\sigma_i^2)b_5P_t,\nonumber\\
a_6&=b_5\sigma_i^2P_t,
\end{align}
and
\begin{align}
&b_1=P_t\|\mathbf{G}\boldsymbol{\tilde{\theta}}\|_2^2,\nonumber
\quad b_2=\sigma_i^2\|\mathbf{F}\mathbf{V}\|_F^2,\nonumber\\
&b_3=N\sigma_i^2\|\mathbf{B}\|^2_F,
\quad b_4=N\sigma_b^2\|\mathbf{A}_s^{-1}\|^2_F,\\
&b_5=16N^2 P_t.\nonumber
\end{align}

Therefore, the optimization problem of PA can be casted as
\begin{align}
\underset{\beta}{\mathrm{min}} \quad &\varepsilon(\beta) \label{Problem}\\
\mathrm{s.t.} \quad &0<\beta<1\tag{\ref{Problem}{a}} .
\end{align}

The derivative of (\ref{varepsilon-beta}) with respect to $\beta$ yields
\begin{align}\label{varepsilon-prime}
\varepsilon^{\prime}(\beta)
=\frac{c_1\beta^4+c_2\beta^3+c_3\beta^2+c_4\beta+c_5}
{(a_4\beta^3+a_5\beta^2+a_6\beta)^2},
\end{align}
where
\begin{align}
&c_1=-a_1a_4,\nonumber
\quad c_2=-2a_2a_4,\nonumber\\
&c_3=a_1a_6-a_2a_5-3a_3a_4,\\
&c_4=-2a_3a_5,\nonumber
\quad c_5=-a_3a_6.\nonumber
\end{align}
Letting (\ref{varepsilon-prime}) equal 0, namely
\begin{align}
\beta^4+d_3\beta^3+d_2\beta^2+d_1\beta+d_0=0,
\end{align}
where $d_3=\frac{c_2}{c_1}$, $d_2=\frac{c_3}{c_1}$, $d_1=\frac{c_4}{c_1}$, and $d_0=\frac{c_5}{c_1}$. Then, we can get 
\begin{align}
\beta^{\text{opt}}
=\underset{\beta\in S_1}{\arg\min}\quad(\ref{varepsilon-beta}),
\end{align}
where $S_1=\{\beta_1, \beta_2, \beta_3, \beta_4\}$ with
\begin{align}\label{optimalbeta}
\beta_{1}& =-\frac14d_3+\frac12R+\frac12P, \nonumber\\
\beta_{2}& =-\frac14d_3+\frac12R-\frac12P, \nonumber\\
\beta_{3}& =-\frac14d_3-\frac12R+\frac12T, \nonumber\\
\beta_{4}& =-\frac14d_3-\frac12R-\frac12T,
\end{align}
where $R=\sqrt{\frac14d_3^2-d_2+y^*}$, and $y^*$ is any real solution of the following equation
\begin{align}\label{y1}
y^3-d_2y^2+\left(d_3d_1-4d_0\right)y
+\left(4d_2d_0-d_1^2-d_3^2d_0\right)=0.
\end{align}
When $R\neq0$, there exists
\begin{align}
P&=\sqrt{\frac{3}{4}d_3^2-R^2-2d_2
+\frac{1}{4}\left(4d_3d_2-8d_1-d_3^3\right)R^{-1}},\nonumber\\
T&=\sqrt{\frac34d_3^2-R^2-2d_2
-\frac{1}{4}\left(4d_3d_2-8d_1-d_3^3\right)R^{-1}}.
\end{align}
When $R=0$, there exists the following 
\begin{align}
P&=\sqrt{\frac{3}{4}d_3^2-2d_2+2\sqrt{y_1^2-4d_0}},\nonumber\\
T&=\sqrt{\frac{3}{4}d_3^2-2d_2-2\sqrt{y_1^2-4d_0}}.
\end{align}

By using Cardan's formula, we can obtain the solutions of (\ref{y1}) as follows
\begin{align}
y_1=&\sqrt[3]{-\frac q2+\sqrt{\left(\frac q2\right)^2+\left(\frac p3\right)^3}}\nonumber\\
&+\sqrt[3]{-\frac q2-\sqrt{\left(\frac q2\right)^2+\left(\frac p3\right)^3}}+\frac{d_2}{3},\nonumber\\
y_2=&\omega\sqrt[3]{-\frac q2+\sqrt{\left(\frac q2\right)^2+\left(\frac p3\right)^3}}\nonumber\\
&+\omega^2\sqrt[3]{-\frac q2-\sqrt{\left(\frac q2\right)^2+\left(\frac p3\right)^3}}+\frac{d_2}{3},\nonumber\\
y_3=&\omega^2\sqrt[3]{-\frac q2+\sqrt{\left(\frac q2\right)^2+\left(\frac p3\right)^3}}\nonumber\\
&+\omega\sqrt[3]{-\frac q2-\sqrt{\left(\frac q2\right)^2+\left(\frac p3\right)^3}}+\frac{d_2}{3},
\end{align}
where
\begin{align}
\omega&=\frac{-1+\sqrt{3}i}2,\nonumber\\
p&=d_3d_1-4d_0-\frac{1}{3}d_2^2,\nonumber\\
q&=4d_2d_0-d_1^2-d_3^2d_0+\frac{1}{3}d_2(d_3d_1-4d_0)-\frac{2}{27}d_2^3.
\end{align}

\subsubsection{When the transmit power at the IoT device is fixed}  there exists an optimal limited reflective power at active IRS. 

In order to obtain the optimal IRS reflect power $P^{\text{opt}}_{\text{IRS}}$, let us define $P_{\text{IRS}}=(1-\beta)P_t$ and $P_{\text{IoT}}=\beta P_t$, then, (\ref{varepsilon-beta}) can be expressed as follows
\begin{align}\label{varepsilon-P-IRS}
\varepsilon(P_{\text{IRS}})
=\frac{e_1 P_{\text{IRS}}^2+e_2 P_{\text{IRS}}+e_3}
{e_4 P_{\text{IRS}}},
\end{align}
where
\begin{align}
e_1&=\sigma_i^2
\|\mathbf{F}\mathbf{V}\|_F^2,\nonumber\\
e_2&=(\sigma_u^2
+N\sigma_i^2\|\mathbf{B}\|^2_F)\cdot(P_{\text{IoT}}
\|\mathbf{G}\boldsymbol{\tilde{\theta}}\|_2^2
+\sigma_i^2),\nonumber\\
e_3&=(P_{\text{IoT}}
\|\mathbf{G}\boldsymbol{\tilde{\theta}}\|_2^2
+\sigma_i^2)^2\cdot N\sigma_u^2\|\mathbf{A}_s^{-1}\|^2_F,\nonumber\\
e_4&=16N^2P_{\text{IoT}}\cdot (P_{\text{IoT}}
\|\mathbf{G}\boldsymbol{\tilde{\theta}}\|_2^2
+\sigma_i^2).
\end{align}
The derivative of (\ref{varepsilon-P-IRS}) with respect to $P_{\text{IRS}}$ yields
\begin{align}\label{varepsilon-prime-P-IRS}
\varepsilon^{\prime}(P_{\text{IRS}})
=\frac{e_1e_4 P^2_{\text{IRS}}-e_3e_4}{(e_4 P_{\text{IRS}})^2},
\end{align}
letting (\ref{varepsilon-prime-P-IRS}) equal 0, we obtain
\begin{align}
P_{\text{IRS}}^{\text{opt}}
=\underset{\beta\in S_2}{\arg\min}\quad(\ref{varepsilon-P-IRS}),
\end{align}
where $S_2=\{P_{\text{IRS}}^{\text{opt1}}, P_{\text{IRS}}^{\text{opt2}}\}$ with
\begin{align}
P_{\text{IRS}}^{\text{opt1}}=
\sqrt{\frac{e_3}{e_1}},\quad
P_{\text{IRS}}^{\text{opt2}}=
-\sqrt{\frac{e_3}{e_1}}.
\end{align}

\section{Proposed DL-based framework for CE}

We initially introduce the fundamental principle of the proposed CDCE algorithm tailored for direct CE. Subsequently, we present the proposed CCCE algorithm. Following that, we demonstrate the application of these two algorithms within the DL-based CE framework of the active IRS-aided uplink IoT system. Lastly, the computational complexity of the proposed two algorithms is provided.

\subsection{Proposed CDCE algorithm}

\begin{figure*}[htpb]
\centering
\includegraphics[width=5.5in]{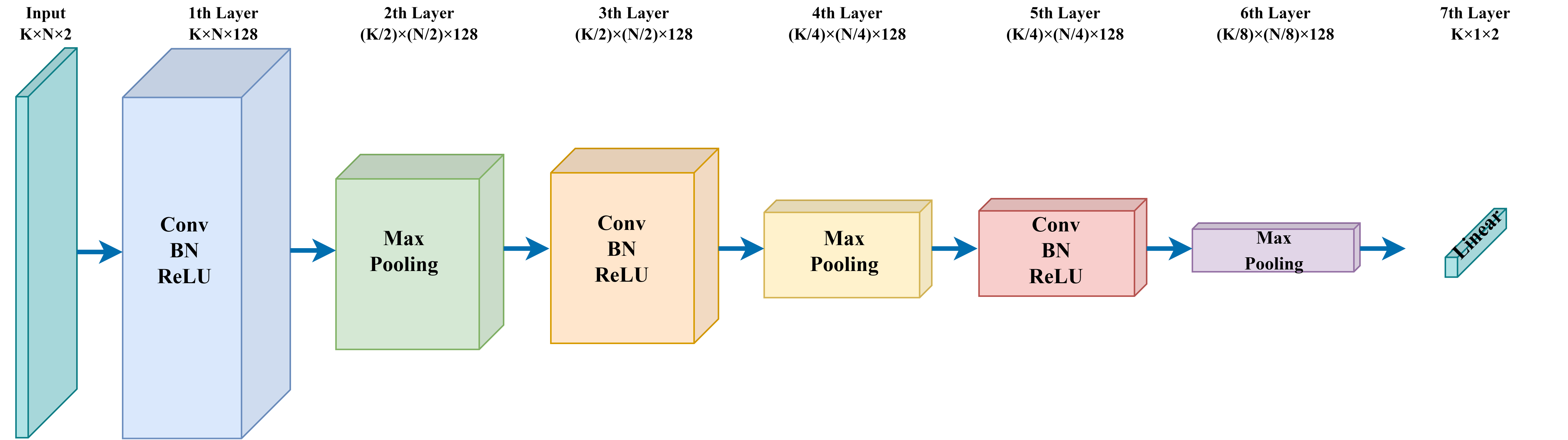}\\
\caption{The proposed CDCE network for direct channel estimation.}\label{Visio-Proposed-CDCE}
\end{figure*}

The proposed CDCE network is implemented on BS side to facilitate the estimation of the uplink DC. Specifically, this network establishes a non-linear function $\boldsymbol{f}_{\boldsymbol{\theta}}$ between $\tilde{\mathbf{Y}}_1$ and $\hat{\mathbf{h}}_{\text{CDCE}}$. The relationship can be mathematically formulated as
\begin{align}
\hat{\mathbf{h}}_{\text{CDCE}}&
=\boldsymbol{f}_{\boldsymbol{\theta}}
(\tilde{\mathbf{Y}}_1),
\end{align}
where $\boldsymbol{\theta}$ stands for the weights.

To effectively train the proposed CDCE network, the BS needs to obtain a sufficient amount of training data in advance. The loss function can be expressed as follows
\begin{align}
\mathcal{L}(\boldsymbol{\theta})
=\frac1{T}\sum_{t=1}^{T}\parallel
\hat{\mathbf{h}}_{\text{CDCE}}-\mathbf{h}_{\text{LS}}\parallel_2^2,
\end{align}
where $T$ represents the size of training dataset. Moreover, the conventional LS CE scheme can be employed to acquire the label $\mathbf{h}_{\text{LS}}$. The goal of training the CDCE network is to minimize $\mathcal{L}(\boldsymbol{\theta})$ by optimizing $\boldsymbol{\theta}$, which is 
\begin{align}
\min_{\boldsymbol{\theta}}\mathcal{L}
(\boldsymbol{\theta})
&=\frac{1}{T}\sum_{t=1}^{T}\parallel
\boldsymbol{f}_{\boldsymbol{\theta}}
(\tilde{\mathbf{Y}}_1)-\mathbf{h}_{\text{LS}}\parallel_{2}^{2}.
\end{align}
When the objective function is determined, the proposed CDCE network undergoes an iterative training process on the training dataset.
In each iteration $i$, $\boldsymbol{\theta}$ are updated via the gradient descent, i.e.,
\begin{align}
\boldsymbol{\theta}_{i+1}
=\boldsymbol{\theta}_i-\alpha_i\mathbf{g}(\boldsymbol{\theta}_i),
\end{align}
where $\mathbf{g}(\boldsymbol{\theta}_i)$ is the gradient vector (GV) for $\boldsymbol{\theta}_i$, and $\alpha_i$ is the learning rate (LR).
After the proposed CDCE network is trained, the BS can directly estimate the DC based on the well-trained CDCE network.

As displayed in Fig.~\ref{Visio-Proposed-CDCE}, the proposed CDCE network consists of three convolution layers, three max pooling layers, and one linear layer.
The spatial features of the channel matrix are exploited by combining the convolution (Conv) operation and the rectified linear unit (ReLU). To enhance the stability of the network and expedite the training speed of the proposed CDCE network, a batch normalization (BN) is introduced between Conv and ReLU.
The 1st, 3rd, and 5th layers are both the ``Conv + BN + ReLU'' operations. As $\tilde{\mathbf{Y}}_1$ is a complex-valued, it is initially crucial to extract the real part and the imaginary part of $\tilde{\mathbf{Y}}_1$ separately as the two input feature maps of the first convolution layer. This operation successfully converts the input of the proposed CDCE network into the real-valued. The second, fourth, and sixth layers are the max pooling layer, where the kernel size and the stride are set to 2. 
By deploying pooling layers to reduce the dimensionality of features extracted from previous layers, the parameter and computational complexity of the next layer are reduced, significantly reducing the consumption of computing resources while ensuring model performance. Additionally, pooling layers prevent overfitting and render the features more robust against noise, enhancing the overall stability and generalization capabilities of the model.
The seventh layer is a linear layer, which maps the output of the previous layer into a vector with $ K\times 1\times2$ elements, where the first $K$ elements denote the real part of $\hat{\mathbf{h}}_{\text{CDCE}}$, and the last $K$ elements correspond to its imaginary part. The hyperparameters of the proposed CDCE network for DC are depicted in Table \ref{tab-CDCE}.

\begin{figure*}[htpb]
\centering
\includegraphics[width=5.5in]{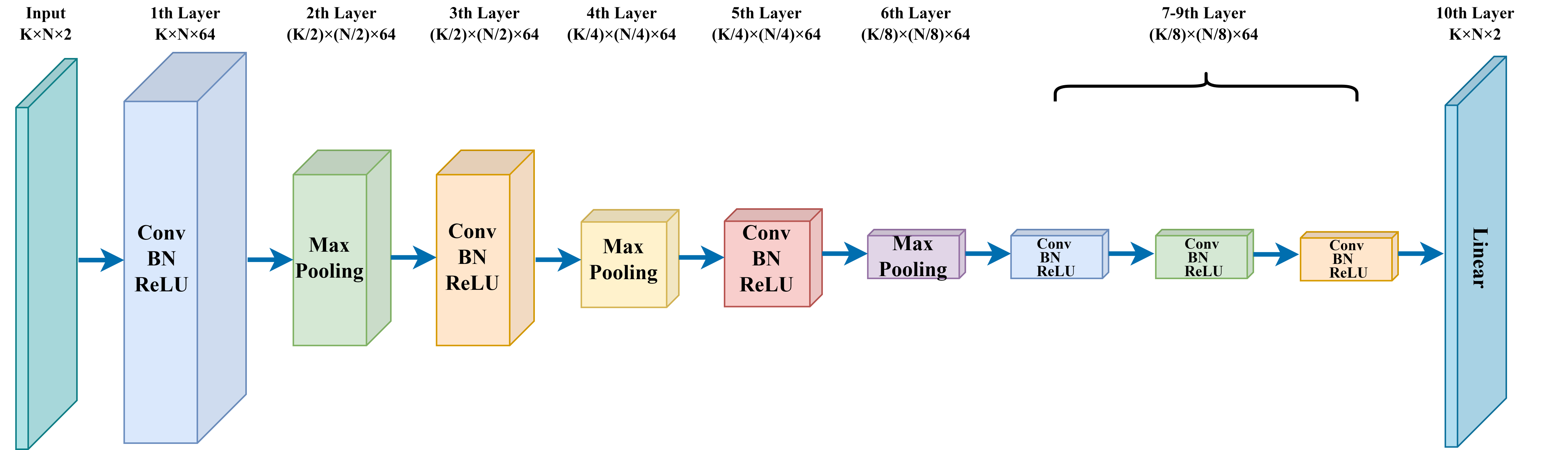}\\
\caption{The proposed CCCE network for cascaded channel estimation.}\label{Visio-Proposed-CCCE}
\end{figure*}

\begin{table}[h]
\begin{center}
\renewcommand{\arraystretch}{1.5}
\tabcolsep=0.37cm
\caption{Hyperparameters of the proposed CDCE network for DC.}
\label{tab-CDCE}
\begin{tabular}{c c c}
\hline
\multicolumn{3}{l}{\textbf{Input}}\\
\multicolumn{3}{l}{The received pilot signal $\tilde{\mathbf{Y}}_1$ with the size of $K \times N$} \\
\hline
\textbf{Layers} & \textbf{Operations} & \textbf{Number of parameters}\\  
1& Conv+BN+ReLU & $(3\times3\times2)\times 128$\\
2 & Max pooling  &-\\ 
3& Conv+BN+ReLU  &$(3\times3\times128)\times 128$\\ 
4 & Max pooling &-\\ 
5 & Conv+BN+ReLU  &$(3\times3\times128)\times 128$\\ 
6 & Max pooling &-\\ 
7& Linear &$(\frac{K}{8}\times\frac{N}{8}\times128)\times(K \times 1\times2)$\\ 
\hline
\multicolumn{3}{l}{\textbf{Output}}\\
\multicolumn{3}{l}{The estimated DC matrix with the size of $K\times 1$} \\
\hline 
\end{tabular}
\end{center}
\end{table}

As presented in \textbf{Algorithm \ref{CDCE-algorithm}}, the proposed CDCE algorithm contains a training part and a testing part, where $i$ is the iteration index and  $I$ denotes the maximum iteration number.

\begin{algorithm} \renewcommand{\algorithmicrequire}{\textbf{Training:}}	\renewcommand{\algorithmicensure}{\textbf{Testing:}}
	\caption{Proposed CDCE algorithm}
	\label{CDCE-algorithm}
    \textbf{Initialization:} initialize trainable parameters, $i=0$, raw training data $\mathbf{h}$, $\mathbf{F}$, $\mathbf{g}$, $\mathbf{V}$, $\mathbf{x}$
	\begin{algorithmic}[1] 
		\REQUIRE 
        \STATE Input direct training set $\tilde{\mathbf{Y}}_1$
        \WHILE{$i<I$}
        \STATE Update $\boldsymbol{\theta}$ by BP algorithm to minimize $\mathcal{L}(\boldsymbol{\theta})$
        \STATE $i=i+1$
        \ENDWHILE
        \STATE Output well-trained CDCE network 
		\ENSURE 
        \STATE Input test set
		\STATE \textbf{do} CE using proposed CDCE network in Fig.~\ref{Visio-Proposed-CDCE}
        \STATE Output estimated DC $\hat{\mathbf{h}}$
	\end{algorithmic}
\end{algorithm}

\subsection{Proposed CCCE algorithm}

Similar to the proposed CDCE network, the proposed CCCE network constructs the non-linear mapping relationship between $\tilde{\mathbf{Y}}_2$ and $\hat{\mathbf{H}}^{biu}_{\text{CCCE}}$, which can be expressed as
\begin{align}
\hat{\mathbf{h}}_{\text{CCCE}}&
=\boldsymbol{f}_{\boldsymbol{\delta}}
(\tilde{\mathbf{Y}}_2),
\end{align}
where $\boldsymbol{\delta}$ is the weights. In addition, $\hat{\mathbf{h}}_{\text{CCCE}}=\text{vec}(\hat{\mathbf{H}}^{biu}_{\text{CCCE}})$ is the vectorization of the CC estimated by the proposed CCCE network. 

Similarly, the loss function is obtained as follows
\begin{align}
\mathcal{L}(\boldsymbol{\delta})
=\frac1{T}\sum_{t=1}^{T}\parallel
\hat{\mathbf{h}}_{\text{CCCE}}
-\mathbf{h}^{biu}_{\text{LS}}\parallel_2^2,
\end{align}
where $\mathbf{h}^{biu}_{\text{LS}}=\text{vec}(\mathbf{H}_{biu})$ can be obtained by equation (\ref{CascadeLS}). The goal of training the proposed CCCE network is as follows
\begin{align}
\min_{\boldsymbol{\delta}}\mathcal{L}
(\boldsymbol{\delta})
&=\frac{1}{T}\sum_{t=1}^{T}\parallel
\boldsymbol{f}_{\boldsymbol{\delta}}
(\tilde{\mathbf{Y}}_2)-\mathbf{h}^{biu}_{\text{LS}}\parallel_{2}^{2},
\end{align}
where the update for $\boldsymbol{\delta}$ are as follows
\begin{align}
\boldsymbol{\delta}_{i+1}
=\boldsymbol{\delta}_i-\beta_i\mathbf{g}
(\boldsymbol{\delta}_i),
\end{align}
where $\mathbf{g}(\boldsymbol{\delta}_i)$ is the GV for $\boldsymbol{\delta}_i$, and $\beta_i$ is the LR.

Unlike the proposed CDCE network, as represented in Fig.~\ref{Visio-Proposed-CCCE}, the proposed CCCE network consists of six convolution layers, three max pooling layers, and one linear layer.
The hyperparameters of the proposed  CCCE network for CC are described in Table \ref{tab-CCCE}.


\begin{table}[h]
\begin{center}
\renewcommand{\arraystretch}{1.5}
\tabcolsep=0.37cm
\caption{Hyperparameters of the proposed CCCE network for CC.}
\label{tab-CCCE}
\begin{tabular}{c c c}
\hline
\multicolumn{3}{l}{\textbf{Input}}\\
\multicolumn{3}{l}{The received pilot signal $\tilde{\mathbf{Y}}_2$ with the size of $K\times N$} \\
\hline
\textbf{Layers} & \textbf{Operations} & \textbf{Number of parameters}\\  
1& Conv+BN+ReLU & $(3\times3\times2)\times 64$\\
2 & Max pooling &-\\ 
3 & Conv+BN+ReLU &$(3\times3\times64)\times 64$\\ 
4& Max pooling &-\\ 
5 & Conv+BN+ReLU &$(3\times3\times64)\times 64$\\ 
6& Max pooling &-\\ 
7 $\sim$ 9 & Conv+BN+ReLU &$(3\times3\times64)\times 64$\\ 
10 & Linear &$(\frac{K}{8}\times\frac{N}{8}\times64)\times(K\times N\times2)$\\ 
\hline
\multicolumn{3}{l}{\textbf{Output}}\\
\multicolumn{3}{l}{The estimated CC matrix with the size of $K\times N$} \\
\hline 
\end{tabular}
\end{center}
\end{table}

\begin{figure*}[htbp]
\centering
\includegraphics[width=5.5in]{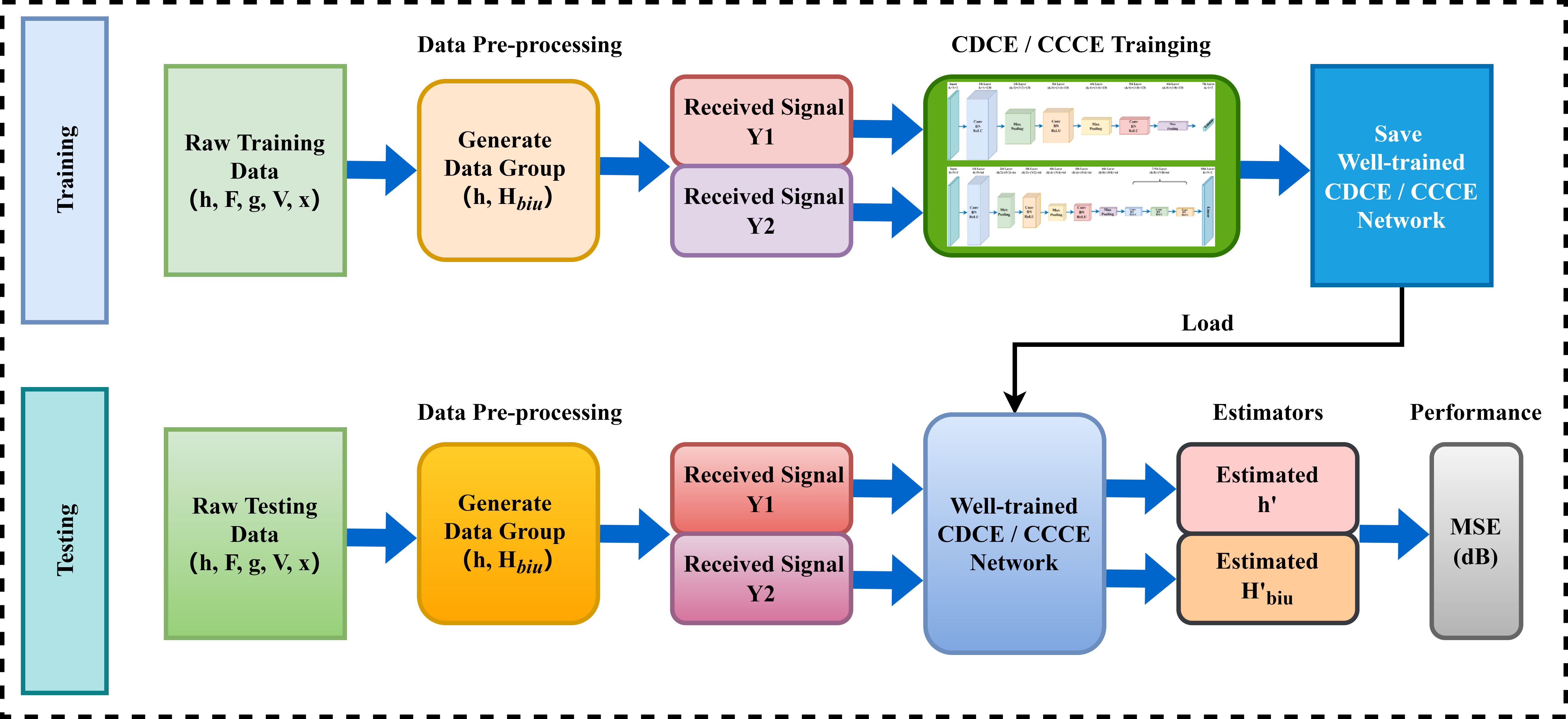}\\
\caption{The proposed DL-based CE framework for the active IRS-aided uplink IoT network.}\label{Visio-zongtitu}
\end{figure*}

The proposed CCCE algorithm for estimating the CC is summarized in \textbf{Algorithm \ref{CCCE-algorithm}}. Based on the designed CDCE network and CCCE network, we apply them to the proposed DL-based framework for CE in the active IRS-aided uplink IoT network, as illustrated in Fig. \ref{Visio-zongtitu}. 

\begin{algorithm} \renewcommand{\algorithmicrequire}{\textbf{Training:}}	\renewcommand{\algorithmicensure}{\textbf{Testing:}}
	\caption{Proposed CCCE algorithm}
	\label{CCCE-algorithm}
    \textbf{Initialization:} initialize trainable parameters, $i=0$, raw training data $\mathbf{h}$, $\mathbf{F}$, $\mathbf{g}$, $\mathbf{V}$, $\mathbf{x}$
	\begin{algorithmic}[1] 
		\REQUIRE 
        \STATE Input cascaded training set $\tilde{\mathbf{Y}}_2$
        \WHILE{$i<I$}
        \STATE Update $\boldsymbol{\delta}$ by BP algorithm to minimize $\mathcal{L}(\boldsymbol{\delta})$
        \STATE $i=i+1$
        \ENDWHILE
        \STATE Output well-trained CCCE network 
		\ENSURE 
        \STATE Input test set
		\STATE \textbf{do} CE using proposed CCCE network in Fig.~\ref{Visio-Proposed-CCCE}
        \STATE Output estimated CC $\hat{\mathbf{H}}_{biu}$
	\end{algorithmic}
\end{algorithm}

\subsection{Computational Complexity Analysis}


For a Conv layer, its computational complexity \cite{DongDeep} can be expressed by
\begin{align}
C_{\text{convolution layer}}=\mathcal{O}(W_1W_2F^2N_{in}N_{out}),
\end{align}
where $W_1$ and $W_2$ are the number of rows and columns of each output feature map, $F$ denotes the side length of the used filter, $N_{in}$ and $N_{out}$ stand for the numbers of input and output feature maps, respectively. 

The computational complexity of a linear layer can be expressed as follows
\begin{align}
C_{\text{linear layer}}=\mathcal{O}(E_1E_2),
\end{align}
where $E_1$ and $E_2$ represent the dimension of input and output, respectively.

Compared to convolutional layers, the computational complexity of a max pooling layer is usually considered negligible \cite{LiuPoolNet}. 

\subsubsection{For the proposed CDCE network}

The computational complexity of the 1st Conv layer is
\begin{align}
C_{d1}=\mathcal{O}(K\cdot N \cdot 3^2\cdot 2\cdot 128)=\mathcal{O}(2304KN).
\end{align}
Similarly, the computational complexity of the 3rd and 5th convolutional layers is 
\begin{align}
C_{d3}=\mathcal{O}(\frac{K}{2}
\cdot\frac{N}{2}\cdot 3^2\cdot 128\cdot 128)=\mathcal{O}(36864KN)
\end{align}
and
\begin{align}
C_{d5}=\mathcal{O}(\frac{K}{4}
\cdot\frac{N}{4}\cdot 3^2\cdot 128\cdot 128)=\mathcal{O}(9216KN),
\end{align}
respectively.
Under the premise of only considering multiplication operations, the computational complexity of the final linear layer is as follows
\begin{align}
C_{d7}=\mathcal{O}(\frac{K}{8}
\cdot\frac{N}{8}\cdot 128\cdot K\cdot 1\cdot 2)=\mathcal{O}(4K^2N).
\end{align}
Therefore, the total computational complexity of the proposed CDCE network is
\begin{align}
C_{d}&=C_{d1}+C_{d3}+C_{d5}+C_{d7}\nonumber\\
&=\mathcal{O}(48384KN+4K^2N).
\end{align}

\subsubsection{For the proposed CCCE network}
Based on Fig.~\ref{Visio-Proposed-CCCE} and Table \ref{tab-CCCE}, the computational complexity of the 1st, 3rd, and 5th Conv layers is 
\begin{align}
C_{c1}=\mathcal{O}(K\cdot N \cdot 3^2\cdot 2\cdot 64)=\mathcal{O}(1152KN),
\end{align} 
\begin{align}
C_{c3}=\mathcal{O}(\frac{K}{2}
\cdot\frac{N}{2}\cdot 3^2\cdot 64\cdot 64)=\mathcal{O}(9216KN),
\end{align}
and
\begin{align}
C_{c5}=\mathcal{O}(\frac{K}{4}
\cdot\frac{N}{4}\cdot 3^2\cdot 64\cdot 64)=\mathcal{O}(2304KN),
\end{align}
respectively.

In addition, the computational complexity of the 7th, 8th, and 9th Conv layer is the same, which is
\begin{align}
C_{c7}=C_{c8}=C_{c9}&=\mathcal{O}(\frac{K}{8}
\cdot\frac{N}{8}\cdot 3^2\cdot 64\cdot 64)\nonumber\\
&=\mathcal{O}(576KN).
\end{align}
When only multiplication operations are considered, the computational complexity of the linear layer of the proposed CCCE network is as follows
\begin{align}
C_{c10}=\mathcal{O}(\frac{K}{8}
\cdot\frac{N}{8}\cdot 64\cdot K\cdot N\cdot 2)=\mathcal{O}(2K^2N^2).
\end{align}
Therefore, the total computational complexity of the proposed CCCE network is
\begin{align}
C_{c}&=C_{c1}+C_{c3}+C_{c5}+C_{c7}+C_{c8}+C_{c9}+C_{c10}\nonumber\\
&=\mathcal{O}(14400KN+2K^2N^2).
\end{align}

\section{Simulation Results}

In this section, first and foremost, the simulation results of the optimal PA strategy capable of minimizing Sum-MSE are presented. Subsequently, the simulation results are provided to validate the performance and effectiveness of the proposed CDCE / CCCE network.

The distances from IoT devices to BS, from IoT devices to active IRS, and from active IRS to BS are 100 m, 70 m, and 60 m, respectively. The path loss at distance $d$ is modeled as follows
\begin{align}
\text{PL}(d)=\text{PL}_0-10\alpha\log_{10}
(\frac{d}{d_0}),
\end{align}
where $\text{PL}_0$ = -30 dB stands for the path loss at a reference distance of $d_0$ = 1m, and $\alpha$ indicates the path loss exponent. Specifically, the path loss exponents of the IoT devices $\to$ BS, IoT devices $\to$ active IRS and active IRS $\to$ BS are set as 3.5, 2.4, and 2.2, respectively. 

Assuming 10000 samples are collected at BS, we proposed that 90\% be allocated to form the training dataset, while the remaining 10\% be designated for the validation dataset.
In addition, we regenerate the 5000 samples used to appraise the performance of the proposed CDCE / CCCE network.
Moreover, the AdamW optimization algorithm is employed to update the model's weight parameters. 
The training process encompasses a total of 200 epochs.
After 200 epochs, we will use the trained network to evaluate its performance.
The LRs of the proposed CDCE and CCCE networks are initialized to 1e-3 and 2.5e-4, respectively, and then reduced to the half of the original level every 35 epochs.



\begin{figure}[h]
\centering
	\includegraphics [width=0.5\textwidth]{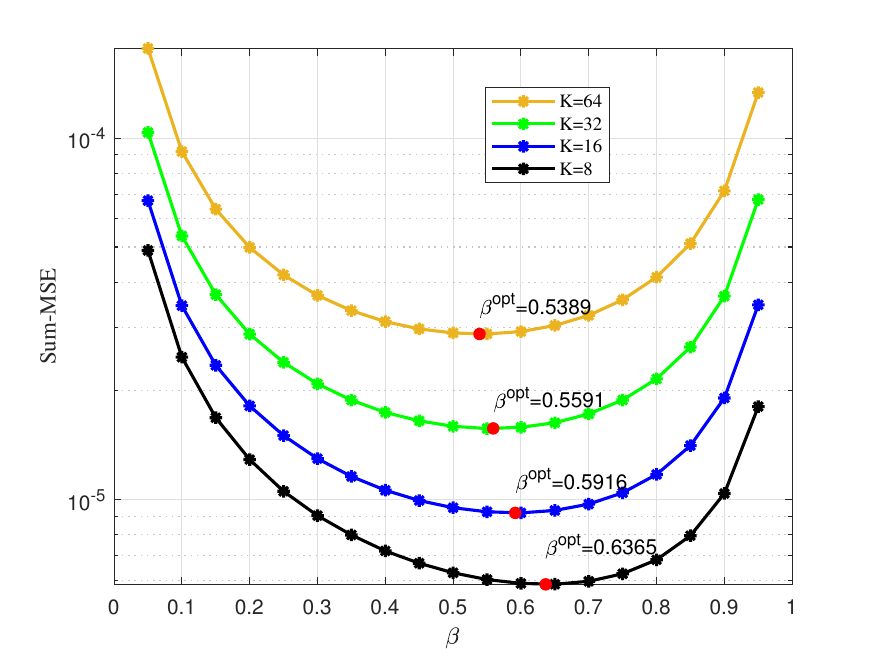}
	\caption{Sum-MSE versus the PAF $\beta$ with different number of BS antennas $K$.}
    \label{x_beta_different_K}
\end{figure}

Fig.~\ref{x_beta_different_K} depicts the Sum-MSE versus the PAF $\beta$ with different $K$. It is evident from the figure that as $K$ increases, the optimal PAF $\beta^{\text{opt}}$ decreases. This indicates that the power allocated to the BS decreases, while the power allocated to the active IRS increases. In addition, as $K$ increases, the value of Sum-MSE also exhibits an upward trend.

\begin{figure}[h]
\centering
	\includegraphics [width=0.5\textwidth]{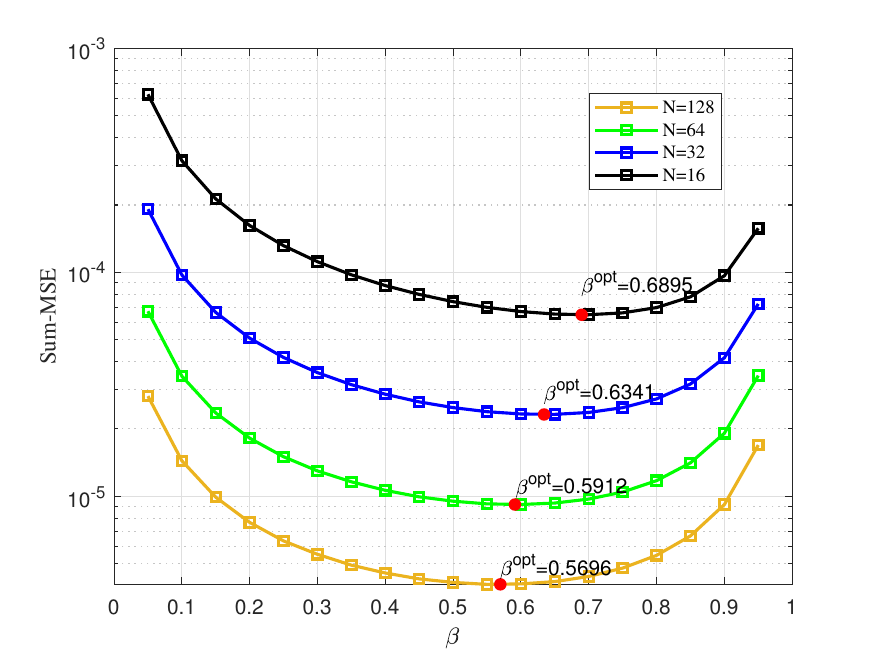}
	\caption{Sum-MSE versus the PAF $\beta$ with different number of active IRS elements $N$.}
    \label{x_beta_different_N}
\end{figure}

Fig.~\ref{x_beta_different_N} illustrates the Sum-MSE versus the PAF $\beta$ with different $N$. It is clear that as $N$ increases, the Sum-MSE decreases, which is consistent with the derived in (\ref{summse}). Furthermore, $\beta$ decreases as $N$ increases. This indicates that a larger proportion of power needs to be allocated to the active IRS.

\begin{figure}[h]
\centering
	\includegraphics [width=0.5\textwidth]{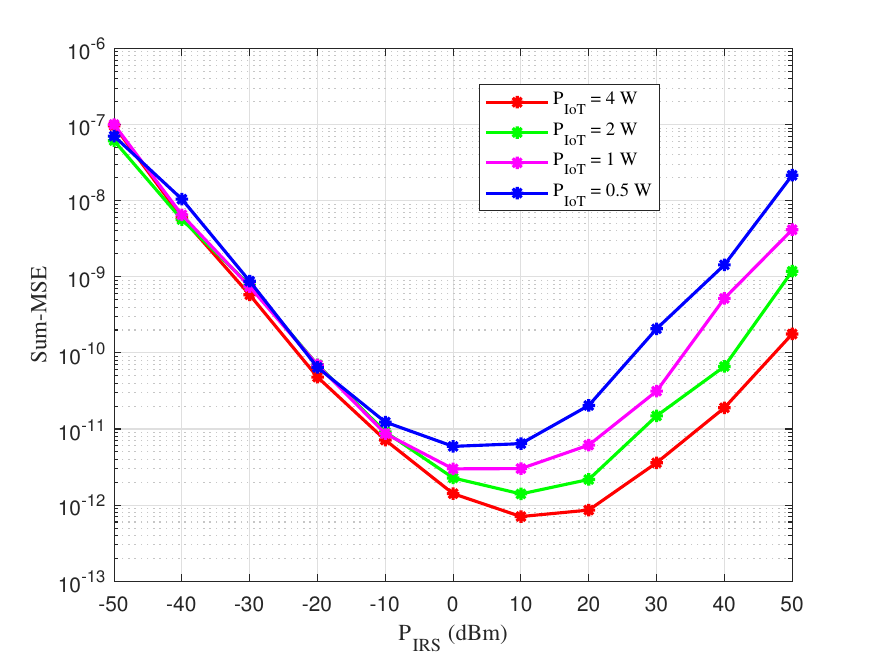}
	\caption{Sum-MSE versus the reflected power $P_{\text{IRS}}$ at active IRS with different transmit power $P_{\text{IoT}}$ at the IoT devices.}
    \label{x_Pi_variable_PBS}
\end{figure}

Figs. \ref{x_beta_different_K} and \ref{x_beta_different_N} display the optimal PA strategy under the total power sum constraint of IoT devices and active IRS. In Fig.~\ref{x_Pi_variable_PBS}, when the transmitted power at the IoT devices is fixed, how the reflected power at the active IRS changes is analyzed. Specifically, Fig.~\ref{x_Pi_variable_PBS} describes the Sum-MSE versus the reflected power $P_{\text{IRS}}$ at active IRS. It is evident that, for a given fixed transmit power at the IoT devices, there exists an optimal reflected power at the active IRS that minimizes the Sum-MSE. Beyond this point, any further increment in the value of $P_{\text{IRS}}$ does not contribute to a reduction in the Sum-MSE. This trend primarily stems from the fact that while the active IRS amplifies signals, it concurrently amplifies noise as well.

\begin{figure}[h]
\centering
	\includegraphics [width=0.5\textwidth]{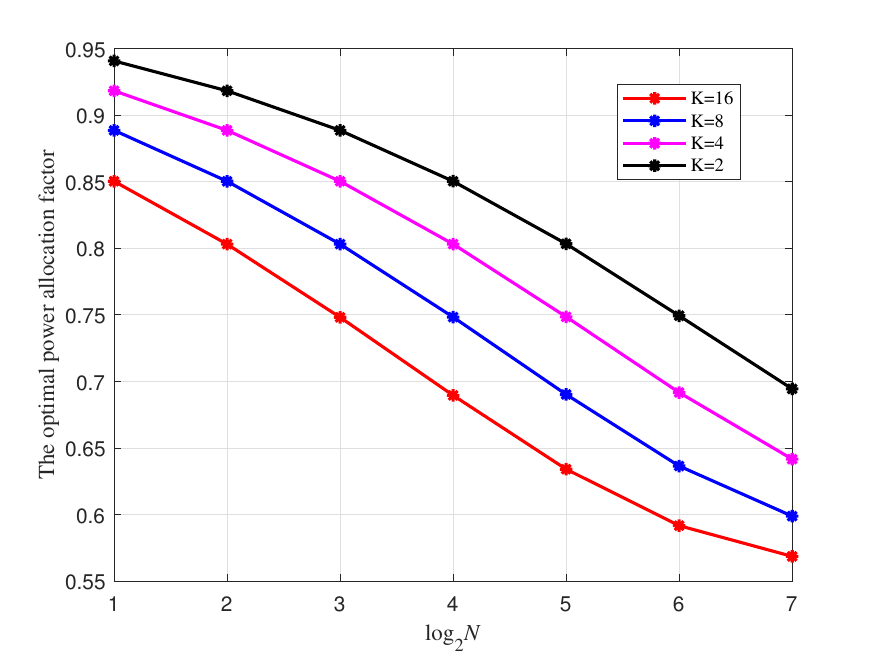}
	\caption{The optimal PAF $\beta^{\text{opt}}$ versus the number of active IRS elements $N$ with different number of BS antennas $K$.}
    \label{x_N_variable_K}
\end{figure}

Fig.~\ref{x_N_variable_K} shows the derived closed-form expression of the optimal PAF in (\ref{optimalbeta}) versus $N$ with different $K$. For a given $N$, the optimal PAF $\beta^{\text{opt}}$ becomes smaller as $K$ increases. This trend aligns with the observations made in Fig.~\ref{x_beta_different_K}. Moreover, when $K$ is held constant, the optimal PAF $\beta^{\text{opt}}$ also exhibits a decreasing trend with an increase in $N$. This particular trend is consistent with the findings presented in Fig.~\ref{x_beta_different_N}.

\begin{figure}[h]
\centering
	\includegraphics [width=0.5\textwidth]{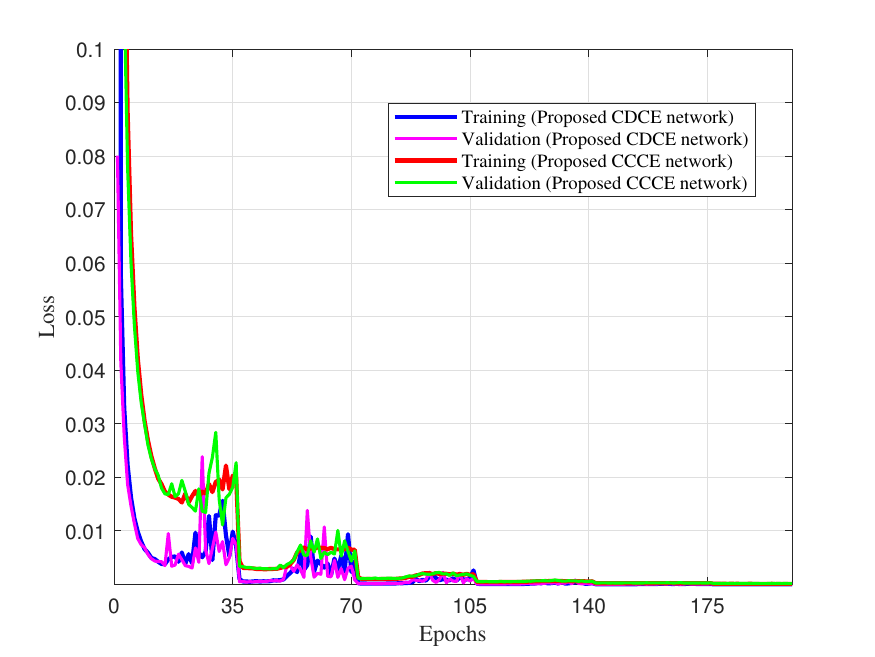}
	\caption{The training and the validation loss of the proposed CDCE and CCCE network versus the training epochs.}
    \label{x_epochs}
\end{figure}

The training and validation performance of the proposed CDCE network and CCCE network are depicted in Fig.~\ref{x_epochs}. We initially adopted a high LR to expedite the model training process and accelerate parameter updates. Subsequently, the LR was decayed by half every 35 epochs to facilitate better convergence of the model. As shown in Fig.~\ref{x_epochs}, as the number of epochs progresses in the training of our proposed model, the corresponding training and validation losses undergo a rapid decline in the initial phase and gradually stabilize in the later stages. Concurrently, our strategy of LR decay effectively contributes to the convergence of the model.

\begin{figure}[h]
\centering
	\includegraphics [width=0.5\textwidth]{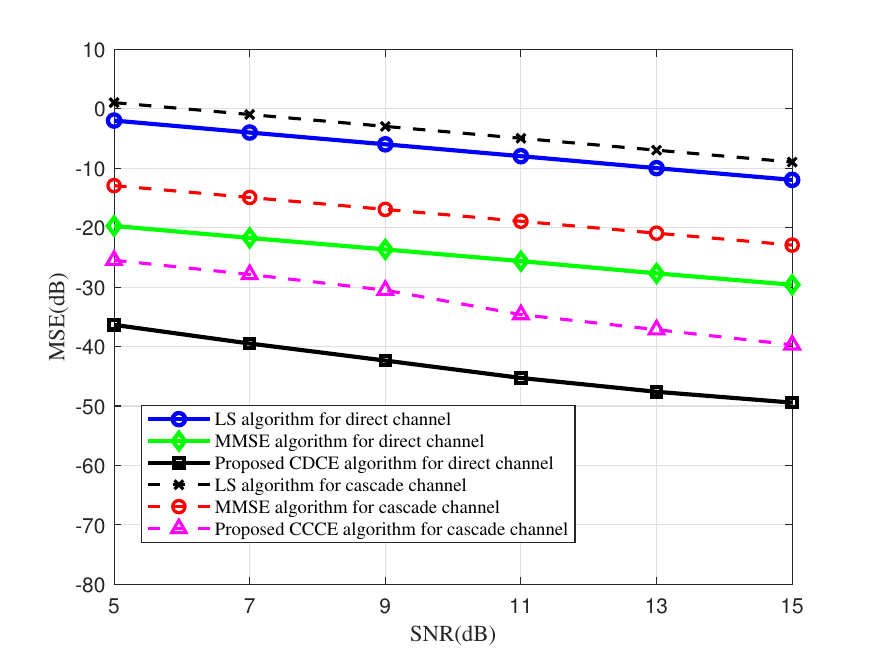}
	\caption{MSE performance versus different SNR under the number of BS antennas $K=16$ and the number of active IRS elements $N=64$.}
    \label{x_SNR_LS_MMSE_DL}
\end{figure}

Fig.~\ref{x_SNR_LS_MMSE_DL} presents the MSE versus SNR with $K=16$ and $N=64$. As evident from Fig.~\ref{x_SNR_LS_MMSE_DL}, the proposed CDCE network outperforms convention LS and MMSE schemes in estimating DCs. Similarly, the proposed CCCE network achieves superior CE accuracy compared to conventional LS and MMSE methods for CCs. It is noteworthy that regardless of the estimation scheme employed, the estimation error for the CC is consistently higher than that for the DC.

\begin{figure}[h]
\centering
	\includegraphics [width=0.5\textwidth]{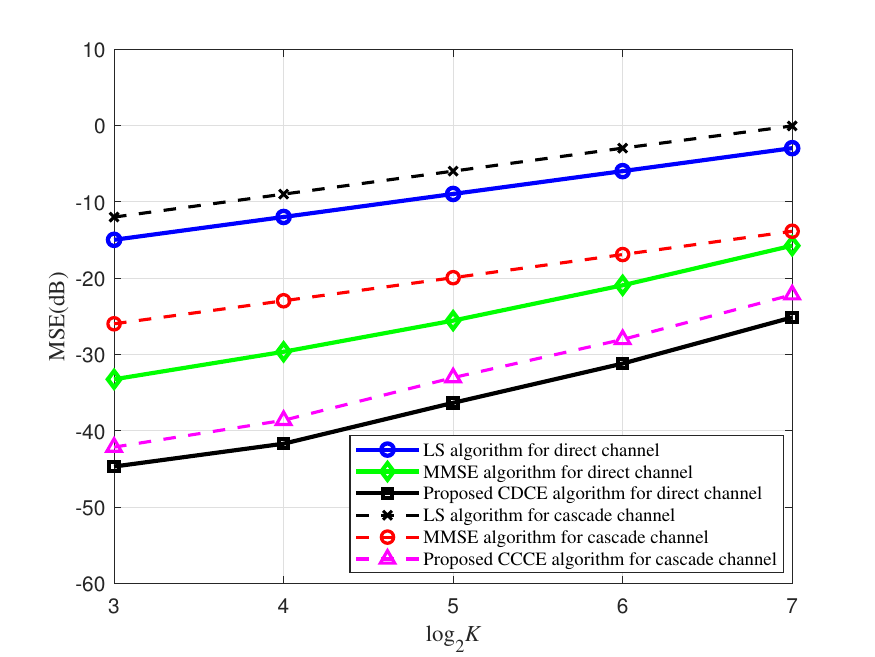}
	\caption{MSE performance versus different number of BS antennas $K$ when the number of active IRS elements $N=128$.}
    \label{x_K_DL_LS_MMSE}
\end{figure}

Fig.~\ref{x_K_DL_LS_MMSE} displays the MSE performance versus different $K$ when $N=128$. As $K$ increases, it is apparent that both the MSE of the DC and the MSE of the CC also increase. This phenomenon is mainly because the size of the estimated channel increases with $K$. Furthermore, the trend observed in Fig.~\ref{x_K_DL_LS_MMSE} aligns with the conclusions drawn from Fig.~\ref{x_beta_different_K}.

\section{Conclusion}

In this paper, power optimization and DL for CE of an active IRS-aided uplink IoT network has been investigated. The LS estimator for direct and cascaded channels was first presented. Subsequently, the corresponding MSE of channel estimator was derived and analyzed. In order to further reduce Sum-MSE, a closed-form expression of the optimal PAF was derived. Furthermore, two channel estimators based on DL were proposed to further improve estimation accuracy. Both of these networks were capable of estimating unknown high-dimensional channels by leveraging known received pilot signals. Simulation results indicated that there exists an optimal PA strategy that minimizes Sum-MSE. Compared with the traditional LS and MMSE benchmark schemes, the proposed CDCE and CCCE algorithms can effectively improve the estimation performance of active IRS assisted uplink IoT systems.

\end{document}